# A Framework for Discovering Optimal Solutions in Photonic Inverse Design

Jagrit Digani[1], Phillip Hon[2], Artur R. Davoyan[3]

[1]*Department of Computer Science, University of California, Los Angeles, CA, 90095*
[2] *NG Next, Northrop Grumman Corporation, Redondo Beach, CA, 90278, USA*
[3]*Department of Mechanical and Aerospace Engineering, University of California, Los Angeles, CA, 90095*
davoyan@ucla.edu

**Abstract**
Photonic inverse design has emerged as an indispensable engineering tool for complex optical systems. In many instances it is important to optimize for both material and geometry configurations, which results in complex non-smooth search spaces with multiple local minima. Finding solutions approaching global optimum may present a computationally intractable task. Here, we develop a framework that allows expediting the search of solutions close to global optimum on complex optimization spaces. We study the way representative black box optimization algorithms work, including genetic algorithm (GA), particle swarm optimization (PSO), simulated annealing (SA), and mesh adaptive direct search (NOMAD). We then propose and utilize a two-step approach that identifies best performance algorithms on arbitrarily complex search spaces. We reveal a connection between the search space complexity and algorithm performance and find that PSO and NOMAD consistently deliver better performance for mixed integer problems encountered in photonic inverse design, particularly with the account of material combinations. Our results differ from a commonly anticipated advantage of GA. Our findings will foster more efficient design of photonic systems with optimal performance.

**Main text**
  Computational design of micro and nanophotonic structures with desired optical responses and dispersion properties has emerged as a versatile engineering tool. A range of inverse design methods [1-8] that allow creating structures with unintuitive properties have been proposed and implemented recently. Such methods enable a breadth of applications from integrated optics [9, 10] and metasurfaces [4, 5, 11-13] to radiative cooling [8, 14] and accelerators on chip [15, 16].
  At the heart of inverse design are numerical optimization algorithms that search across a virtual space of possible geometries and materials to find designs with responses close to desired target specifications. However, in many practical scenarios the search space becomes excessively large and complex, such that it is computationally intractable. In this case numerical optimization may land at a local optimum – a best guess on a small subspace of entire search space. At the same time many emerging photonic applications necessitate design of structures that operate close to their performance and efficiency limits. Examples of such structures may include ultralow loss resonators and waveguides for quantum information processing [17-21], ultralight-weight high-power metasurfaces for propulsion [22, 23], near-unity broad-band antireflection coatings for displays and solar cells [24], and ultra-high efficiency structures for concentrator photovoltaics [25-27], to name few. Such systems necessitate methods that can discover photonic designs close to the global optimum (i.e., best possible solution of the entire search space). Notably, in an overwhelming number of design problems *a priori* knowledge of performance limits (i.e., the bounds on the global optimum) are missing, which further complicates the global optimum search. Indeed, as the search target is unknown it becomes unclear when to terminate the algorithm, particularly in the case of computationally expensive objective functions. In addition, many photonic problems with discrete set of geometric constraints and material options feature search spaces with objectives functions that are non-smooth with very abrupt variations [8, 28]. Identifying global optimal solutions in such cases is a nontrivial task. This is where a framework that may enable a more systematic way of finding the best possible solutions in a finite amount of time is necessary.

Here, taking a simple multimaterial multilayer stack as a model problem (Figure 1a), we study a framework that allows establishing optimization strategies suitable to finding best possible solutions for a given objective function. Our study reveals that properly adjusted direct search methods, such as NOMAD, consistently deliver better performance on complex search spaces. We further observe that optimization on structured search spaces allows faster and more reliable discovery of best possible solutions. Our results indicate strategies for defining complex photonic optimization problems and outline pathways for efficiently implementing optimization algorithms.

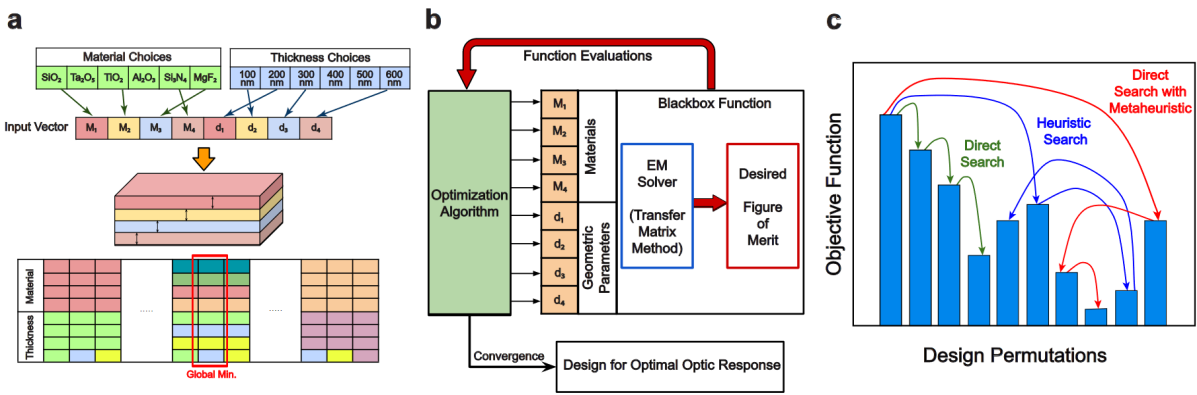

*Figure 1: (a) Schematic representation of a multilayer stack optimization problem. Here a discrete set of materials and thicknesses constitute a mixed integer search space. Each element of a search-space is a multidimensional vector. Global optimization methods traversing through possible vectors of the search space aim to discover one that yields an optimal performance for a given figure of merit. (b) Block-diagram representation of a respective blackbox optimization process. Here optimization algorithms operate on a space of multidimensional vectors treating them as input variables of an objective function. (c) Schematic illustration showing the difference between global and local optimization and displaying search strategies employed by direct search and heuristic optimization algorithms.*

**Results and Discussion**

A general optimization problem for photonic inverse design can be formulated as finding the minimum of a given objective function $f$ on a space of materials and geometries: $\min_{m \in M, d \in D} f(m, d)$, where $f(m, d)$ is the objective function that returns the figure of merit (FoM) of an optical system examined, and $M$ and $D$ constitute the *search-space* of material and geometric constraints that describes the optimization problem (Figure 1a). Global optimization methods attempt to find the global optimum of an objective function $f$ as opposed to stopping on the discovery of a local optimum [29]. In many cases the difference between an arbitrary local minimum and the global minimum could mean a considerable performance loss. However, global optimization is significantly more challenging since there is no guaranteed way to find the global optimum without examining the entire search-space (i.e., traversing all combinations on $M$ and $D$), which is often not feasible. Complex and non-smooth (i.e., discrete) objective functions frequently arising in photonic design problems, further lead to an increased number of local minima [8, 28], which makes discovery of the global optimum even more challenging. Searching through such complex non-smooth mixed-integer spaces suggests treating the objective function, $f$, as a blackbox, which deterministically maps input parameters onto system's responses without detailing the inner-workings of the problem (Figure 1b). This quality makes blackbox optimization useful for studying computationally-intensive non-smooth objective functions where gradients may not be well defined.

Blackbox optimization methods can be divided into two classes [29]: heuristic and non-heuristic methods, with a difference being that while heuristic methods work in practice, there is no mathematical guarantee of them successfully converging to the global optimum [29]. Optimization methods such as the Genetic Algorithm [30] are classified as heuristic methods, since their search strategies are to a certain extent based on chance but, nonetheless, they are effective in practice [29], Figure 1c. Non-heuristic methods employ stringer search mechanics that depends on the objective function and a more mathematically rigorous convergence analysis [29]. Examples of non-heuristic algorithms include Direct Search methods [31] (e.g., NOMAD). However, such direct search methods are more likely to converge to a local minimum, Figure 1c. Heuristic methods, due to their more random search strategies, may take longer to discover the global optimum, but they are usually able to escape local minima effectively as compared to some non-heuristic methods (Figure 1c). Combination of direct search strategies with metaheuristic frameworks may offer a more robust pathway for discovering a global optimum (Figure 1c). Choice of an algorithm may become a key consideration when searching for solutions near global optimum. While there have been many studies of fixed-cost (i.e., best result in a given number of steps) analysis of the efficiency of different algorithms [8, 32], it remains unclear when a given set of algorithms should be implemented and how efficient they are in escaping local optimum and converging to a global optimum while optimizing both material and geometric parameters. As there is no guaranteed way of knowing that the global optimum has been discovered until the entire search space is evaluated, an algorithm's ability to consistently get reasonably close to, if not directly discover the global optimum in a short timeframe becomes a major design criterion. To the best of our knowledge at present there is no unified approach that allows examining algorithm reliability in finding solutions approaching global optimum, particularly for complex mixed-integer optimization problems.

For the sake of concept demonstration, we reduce a generic optimization problem to a study of multilayers made of different material combinations (Figure 1a). While simple to implement, optical multilayers provide an attractive platform to understand physical and computational processes at hand. Furthermore, despite the seeming simplicity, inverse design of multilayer stacks belongs to a class of mixed-integer optimization problems [8, 28, 33] that do not have a straightforward solution. In addition, as we will show below, complex FoMs may be introduced, which may result in search spaces with many randomly distributed local minima. For a $n$-layer multilayer structure the optimization problem reduces to finding $\min_{m \in M, d \in D} f(m, d)$, where $M \subset \mathbb{Z}^n$ is the set of material choices and $D \subset \mathbb{R}_+^n$ is the set of possible layer thicknesses. This search space is schematically represented in Figure 1a. With the growth of the possible choices of materials and geometries the search space size grows exponentially making even this physically simple problem computationally intractable (in the order of $\sim(|M||D|)^n$, if $D$ is discrete and finite). We note that optimizing the design of a multilayer structure has been studied before as a Mixed Integer Optimization problem [8], and an approach for implementing a memetic algorithm (an evolutionary algorithm similar to GA) was proposed. Here we propose and study a more generic framework aimed at understanding global search of diverse blackbox optimization algorithms, including GA.

To understand the relation between the objective function for a given FoM and algorithm's efficiency and reliability in discovering global optimum, we propose a two-step framework involving "*fixed-target*" and "*fixed-cost*" analysis [34]. Our first step is based on approximating a large computationally intractable search space with a small subset that can be fully computed in a reasonable amount of time – the "*known-subset*" of the search space. For this purpose, we restrict the number of materials, layers, and thickness choices to keep the known-subset to a reasonable size and compute the objective function at each point. This allows finding rigorously a global optimum of the known-subset. We then tune and gauge the efficiency and reliability of an algorithm to find the global optimum on a known-subset (i.e., a fixed-target analysis is performed). At a next step we consider the entire search space (where global optimum is not known) and validate the efficiency of tuned algorithms in advancing through a search space in a fixed number of blackbox function calls (i.e., fixed-cost analysis is performed). In this step we allow the thicknesses to be continuous variables instead of discrete choices and further allow for the entire set of materials and a larger number of layers to reflect the application-based use-case of such optimization strategies. Such a two-step approach is reminiscent of bootstrapping techniques used widely in statistics and machine learning [35].

In our fixed-target analysis step, we track the number of function evaluations it takes the algorithm to find the global optimum on the smaller known-subset. To quantify the measure of success in discovering a global optimum we introduce a probability of success:

$$P_{success}(N) = \lim_{Q \to T} \frac{S(n_i \leq N)}{Q} \quad (1)$$

where $T$ is the size of the known-subset (i.e., all possible material and geometry combinations of a smaller subset of an entire search space), $Q$ is the number of trial runs (each with randomly produced starting point), $S(n_i \leq N)$ is the number of trials in which global optimum was discovered before reaching $N$ evaluations. To be specific, consider a known-subset of $T$ elements, $\{f_1, ..., f_T\}$. We then randomly initiate a starting point, an initial condition, $f_q$, and trace algorithm convergence to the global optimum. Depending on the initial condition, $f_q$, an algorithm requires $n_q$ function evaluations to discover the global optimum. Note that in practice, as the search space is not known, choice of initial conditions is indeed random (a best starting guess), which further justifies our approach. For $Q$ distinct initial conditions (ideally $Q \to T$) we obtain a set $\{n_1, ..., n_Q\}$ which provides a distribution of function evaluations needed to find a global optimum for randomly initiated starting points. Then for a given number $N$ ($N \leq T$) we identify evaluations $n_i$ that resulted in success before reaching $N$ (i.e., $n_i \leq N$); $S$ is then the size of the set $\{n_i : n_i \leq N\}$. Probability is then found as $P = S/N$. Such a definition of a measure of success allows comparing performance of different algorithms on arbitrarily complex optimization problems (i.e., search spaces described with arbitrarily complex objective functions).

To illustrate our framework below we study several conceptually different model problems that also have direct practical applications. We consider the following blackbox optimization algorithms, which are representative of heuristic and direct search methods: Nonlinear Mesh Adaptive Direct Search (NOMAD) [36-39], Simulated Annealing (SA) [40-43], Particle Swarm Optimization (PSO) [44-46], and the Genetic Algorithm (GA) [47-50]. Each of these algorithms is widely used in a diverse range of photonic optimization problems. However, we stress once again that there is no framework that allows assessing performance of these algorithms in finding global optimum while optimizing for material and geometric parameters.

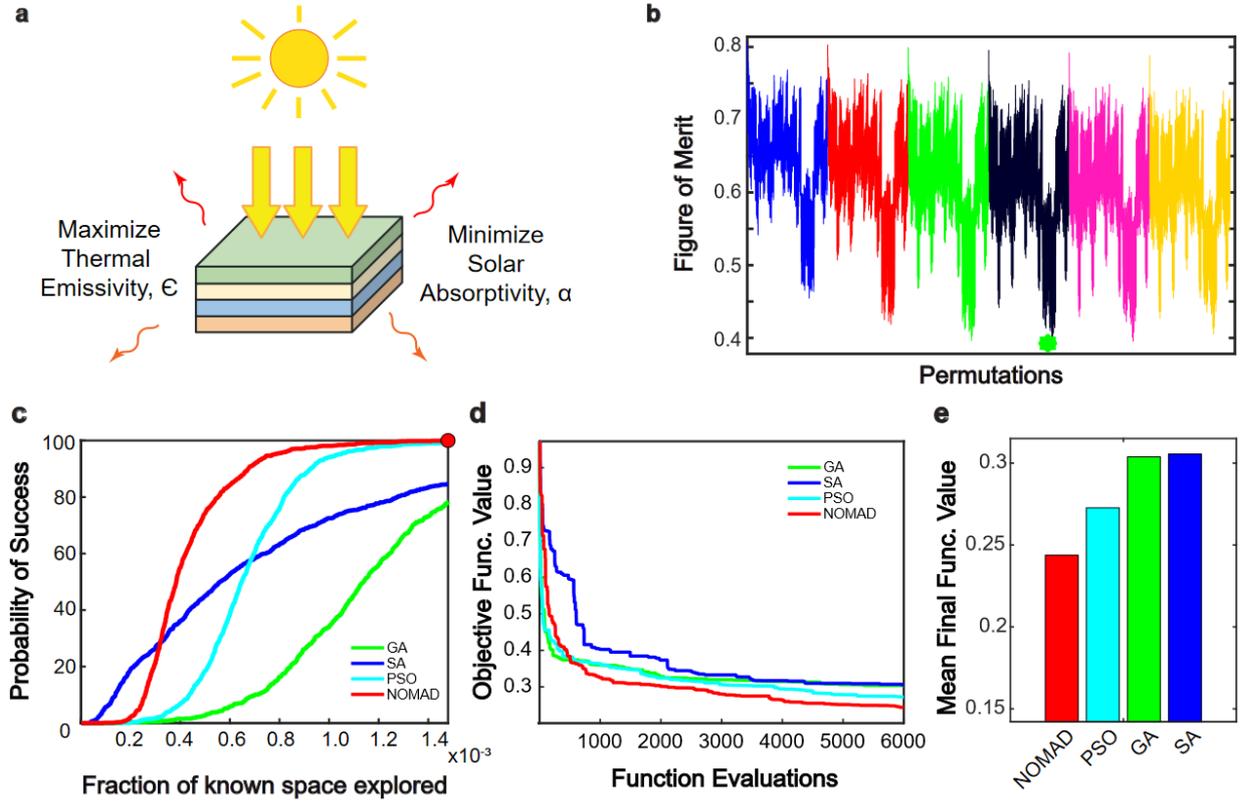

*Figure 2. (a) Schematic Illustration of the optimization problem. Optimization objective: minimize solar absorptivity and maximize emissivity. (b) One-dimensional graph of the objective function designed on the 8-dimentsional known-subset. A small fraction of the permutations on the known-subset is shown. Each point corresponds to a different multilayer structure, whereas different colors represent different material permutations. A marker point denotes a global optimum of the known-subset. (c) Probability of success in finding the global optimum for fixed-target search on a known-subset for different algorithms. (d) Fixed-cost performance analysis. Average best FoM found by the algorithms is shown. (e) Bar chart of the average best FoM found by the algorithms after 6000 objective function evaluations.*

The first optimization problem we consider is a common problem encountered in solar spectrum management, illumination, and radiative cooling [51-55]. In particular, such systems require optimizing for broadband optical responses that cover both visible and infrared spectral bands. For the sake of concept demonstration consider a simple problem of designing a structure with minimal solar absorbance, $\alpha \to 0$, and a maximized thermal emissivity, $\epsilon \to 1$ (Figure 2a). The optimal multilayer structure according to this FoM would not heat up with ease and could ideally be applied as coating to maintain temperatures of a structure out in the sun [56]. Such a structure may find use in design of spacecraft coatings for thermal management [54, 57]. As an example we define the objective function simply as:

$$f = \alpha_{sun} + (1 - \epsilon) \quad (2)$$

where $\alpha_{sun}$ is the normal solar absorptivity of the multilayer structure and $\epsilon$ is its blackbody emissivity, given by $\alpha_{sun} = \int_{\lambda_{min}}^{\lambda_{max}} \alpha_\lambda(\lambda) I_{sun}(\lambda) d\lambda / \int_{\lambda_{min}}^{\lambda_{max}} I_{sun}(\lambda) d\lambda$ and $\epsilon = \int_{\lambda_{min}}^{\lambda_{max}} \epsilon_\lambda(\lambda) I_{BB}(\lambda, T) d\lambda / \int_{\lambda_{min}}^{\lambda_{max}} I_{BB}(\lambda, T) d\lambda$, respectively; $I_{sun}(\lambda)$ is the spectral power density of the solar radiation, and $I_{BB}(\lambda, T)$ is the spectral power density of a black-body at temperature $T$. $\alpha_\lambda(\lambda)$ and $\epsilon_\lambda(\lambda)$ are the spectral absorptance and emittance, respectively, which in our model problem are obtained using a transfer matrix method [58] (see also Methods). We assume that the structure is in thermal equilibrium at room temperature of $T = 298K$ and optimize optical responses in a spectral band from $\lambda_{min} = 0.3\ \mu m$ to $\lambda_{max} = 14\ \mu m$. In our study we consider a 15-layer multilayer structure, where each layer is selected from 15 possible material choices, and assume that each layer thickness may vary between 100 nm and 1000 nm. Such choice of parameters results in a massive search space:

$$M = \{MgF_2, Al_2O_3, HfO_2, SiO_2, AlN, TiO_2, Ta_2O_5, Si_3N_4, SiC, Ti, Cr, Cu, Ag, Au, Al\} \quad (3.1)$$
$$D = [100nm, 1000nm] \quad (3.2)$$

which yields in the order of $\sim 15^{15}$ possible material permutations alone. Obviously, such a problem statement results in a search space that cannot be computed in a reasonable time with a regular computer. We stress that the aim of our analysis is to understand a framework for finding a possible solution on a complex search space; as such finding a best possible structure for a given application is not our objective (in particular, in case a best performance is sought for a more informed selection of materials is needed). Hence, we deliberately introduce metals to assess how algorithms will perform with materials that exhibit drastically different optical responses.

Following our two-step approach, we first select a subset of the entire search space to make the known-subset. Our known-subset is a 4-layer structure defined on a subspace of material and discrete layer thickness choices (selected rather randomly):

$$M' = \{SiO_2, Ta_2O_5, TiO_2, Al_2O_3, Si_3N_4, AlN\} \quad (4.1)$$
$$D' = \{100nm, 200nm, 300nm, 400nm, 500nm, 600nm\} \quad (4.2)$$

Each layer is then represented by a pair of numbers $(i, j)$, where $i$ denotes a material index $m_i \in M'$ and $j$ denotes a layer thickness index $d_j \in D'$, respectively. This subset contains only 1.68 million possibilities, all of which can be computed fairly quickly and the global optimum can be found directly (by traversing the entire search space). It is easy to show that the search space is 8 dimensional, as each possible 4-layer structure design is represented by an 8 element vector (a material and thickness for each layer), see also Figure 1a. Figure 2b displays the objective function plotted for a small section of the known-subset of 4-layer structures resolved on a 2D plot. Different colors represent several different material permutations. Importantly, the objective function exhibits some structure (also visually seen) across permutation of adjacent materials and geometries. As such, we expect that all the algorithms will perform relatively well in optimizing such FoM.

For this known-subset we calculate the probability of finding the known global optimum for four different optimization algorithms (i.e., GA, PSO, NOMAD and SA), Figure 2c. Specifically, for each algorithm we randomly initiate 1000 starting points and examine algorithm convergence. Probability is then calculated according to Eq. (1). As an example, NOMAD's probability is ~80% at 0.06% of the known-subset explored, which corresponds to almost 1000 blackbox function calls. This means that NOMAD had successfully found the global optimum of the objective function by the time it had made 1000 function evaluations in 80% of randomly selected starting points. Furthermore, NOMAD shows faster and more reliable performance for this objective function as compared to other algorithms. In particular, NOMAD has successfully found the global optimum of the known-subset for all 1000 random initial conditions. PSO also shows a good performance in terms of efficiency, but it does fail to find the global optimum in a number of trials. That is, starting from some initial conditions results in PSO converging to a local minimum without discovery of a global optimum. Nonetheless, PSO performance in optimizing this objective function is close to that of NOMAD. The Genetic Algorithm, in contrast, discovers the global optimum in all trials, but it requires a higher evaluation cost to do so. Its ability to converge to a global optimum is expected since GA's search mechanism makes it easier for escaping local minima, Figure 1c. However, the embedded randomness (i.e., heuristic) in traversing the search space makes it less efficient as compared to direct search methods that make use of the search space structure (Figure 2b). Simulated Annealing performs similarly to genetic algorithm, but it has a higher tendency of getting stuck at local minima and not finding the actual global minimum.

According to this analysis, the fixed-target analysis on the known-subset predicts that NOMAD is optimally suited to optimizing relatively structured high-dimensional objective functions. We expect NOMAD to converge faster than purely heuristic algorithms on such structured spaces. To verify this hypothesis we now expand our search to the full search space. In Figure 2d we plot the average best-score for a fixed-cost analysis assuming 6000 total function evaluations (in our fixed cost analysis to expedite computations only 10 random initial conditions are assumed for each algorithm). The performance of algorithms on a much larger search space is reminiscent of that on the smaller known-subset, which validates our hypothesis. A somewhat similar inference is encountered in statistics and machine learning, where smaller data sets (bootstrap or training set) are used to gain knowledge about the larger and more general dataset [35]. We note that a mathematically rigorous proof such inference (i.e., mapping the dynamics of known-spaces onto larger generic search spaces sharing defined for the same FoM) does not exist. However, our computations study (not presented here for the sake of brevity) of 3 and 5 layer structures with different combinations of materials and thickness choices (Eqs. (4.1) and (4.2)) further verifies that such inference can be made and that a framework we propose is justified and valid. In particular, we observe that NOMAD and PSO outperform GA and SA which take >175% more evaluations on average. Both NOMAD and PSO algorithms benefit from their search mechanics exploiting the structure of the objective function (Figure 2b). GA and SA take longer to explore the search-space and find well performing points since their search mechanics is probabilistic and does not exploit the structure of the objective function. Finally we note that NOMAD exhibits the best performance for this objective function as is also expected from our known-subset analysis.

Next, we study more "chaotic" and less structured objective functions, where we expect that algorithms' performance may change. To examine such a possibility, we consider an optimization problem described by a "digital spectrum" model (Figure 3a) [8]. Specifically, we search for multilayer structures that exhibit optical responses maximally close to a target digital spectrum. These digital spectra are obtained by setting the reflectance and absorptance to 1 in randomly selected intervals and to 0 outside of these intervals (Figure 3b). The FoM is then founds as a mean square error between these target digital spectra and the actual spectral reflectance and absorptance:

$$f = \frac{1}{N} \sum_{\lambda=\lambda_{min}}^{\lambda_{max}} \left(\alpha_{target}(\lambda) - \alpha(\lambda)\right)^2 + \frac{1}{N} \sum_{\lambda=\lambda_{min}}^{\lambda_{max}} \left(R_{target}(\lambda) - R(\lambda)\right)^2 \qquad (5)$$

where $\alpha_{target}$ and $R_{target}$ are digital spectral target absorptance and reflectance, respectively; and $N$ is the number of points in the spectrum of analysis (see Methods).

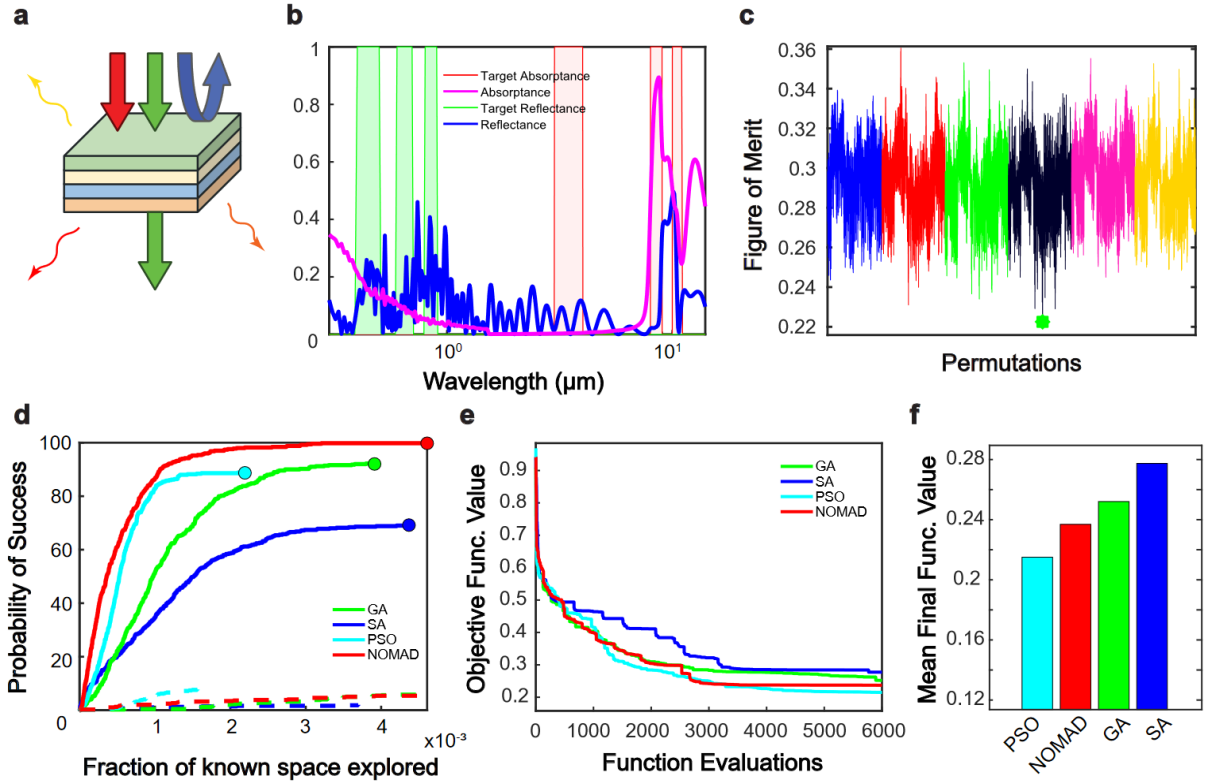

*Figure 3: (a) Schematic Illustration of the optimization problem. The objective is to find the structure with a performance closest to the target digital spectrum. (b) Schematic illustration of the target digital spectrum, and reflectance and absorptance spectra closest to the digital spectrum, found by NOMAD after 6000 function evaluations. (c) One-dimensional graph of the objective function designed on the 8-dimentsional known-subset. A small fraction of the permutations on the known-subset is shown. Each point corresponds to a different multilayer structure, whereas different colors represent different material permutations. A marker point denotes a global optimum of the known-subset. (d) Probabilities of success in finding the global optimum (dashed curves) and solutions within 5% of the global optimum (solid curves) for fixed-target search on a known-subset for different algorithms. The circle markers denote algorithms reaching their respective stopping criteria. (e) Fixed-cost performance analysis. Average best FoM found by the algorithms is shown. (f) Bar-chart depicts the average best FoM found by the algorithms after 6000 function evaluations.*

A section of the known-subset is plotted in Figure 3c. It is clear that the objective function in this case is much more irregular, as compared to the previous example (Figure 2b) (note that a seeming periodicity in Figures 3c, as well as in Figure 2b, arises from reduction of the 8 dimensional space Eq. (4) onto 1D graphical representation in permutation space). The objective function features a higher frequency of local minima and does not exhibit strong patterns that algorithms' internal search mechanism could potentially exploit. We note that our notion of "regularity" is rather ill defined. While treating the objective function resolved on a 1D permutation space (see e.g., Figure 3c) as a probability distribution allows implementing methods of statistical analysis to measure the degree of regularity (e.g., by calculating statistical entropy), we are not aware of methods allowing to analyze multidimensional spaces on which algorithms are operating (e.g., 8 dimensions in the known-subset case studied here). To check whether more irregular objective functions make discovery of optimal solutions harder we plot the probability of finding a global optimum of the known-subset (Eq. (4)). We find that all algorithms perform rather poorly. The global minimum is discovered in a very few of the trial runs, as shown by the dotted curves in Figure 3d. Observing the performance of the algorithms in reaching within 5% of the global minimum (solid lines in Figure 3d) reveals that solutions close to a global minimum are consistently discovered. We therefore conclude that algorithms when searching through such irregular objective functions get stuck in local minima. Indeed, in a limiting case of a search space represented by a uniform random variable distribution, identifying a global optimum would require traversing through all possibilities. Notably even on such highly irregular search spaces NOMAD performs rather efficiently in getting within 5% of the global minimum, which is a strong result in its favor.

Our fixed-cost analysis on the full search space, Figure 3e, shows that all algorithms perform similarly, which is expected due to an irregularity of the search space. By 6000 function evaluations all algorithms come to close results in the average best FoM found (averaged over results obtained for 10 random initial conditions). NOMAD and PSO perform slightly better than GA and SA. While best objective function value is discovered by NOMAD (plotted in Figure 3b), PSO is more consistent in finding better scores for randomly initiated starting points, see Figures 3e and 3f where average FoM is plotted. These results also demonstrate that all algorithms are well optimized to escape local minima even for such high irregular objective functions.

As our last example, we consider a more elaborate optimization problem. In particular, we study a design of a lightsail for laser-driven propulsion [22, 23]. Unlike Breakthrough Starshot mission [23] here we consider more relaxed conditions suitable for near-term missions. Specifically, we assume that a laser source with 1MW power is incident upon a 10cm$^2$ lightsail. To ensure optimal propulsion efficiency we search for lightsails with a smallest possible mass, $m$, which at the same time yield high reflectivity in the 1050nm -1080 nm spectral window. In addition, we require that the temperature of the entire structure is kept below 1000K. Such a problem requires a complex multiobjective optimization that takes into account mass, temperature and optical properties into account. For this purpose we introduce the following multiobjective function:

$$f = \frac{m}{m_{max}} + \exp\left(\frac{T}{1000K}\right) + \frac{1}{N}\sum_{\lambda=1050nm}^{1080nm}\left(1 - R(\lambda)\right)^2 \qquad (6)$$

here $m_{max}$ is the maximum possible mass of a structure given the possible choices, $T$ is the temperature of the structure, $R$ is its spectral reflectance, and $N$ is the number of points in the spectrum of analysis. The temperature is found from radiation balance between laser power absorbed and power emitted as thermal radiation at temperature $T$. Here we choose reflectance and mass weights in this FoM bound within 0 and 1 range, while temperature contribution is strongly penalized for designs with temperatures over $T > 1000K$. Note that here again for the sake of understanding algorithm performance on complex search spaces we assume a bank of 15 material, Eq. (3.1), many of which (e.g., metals) are clearly not suitable for laser propulsion [23]. Design of a high performance structure while uses of a similar objective function will require a more careful preselection of materials. We stress once again that our aim is to explore the optimization framework rather than identify best possible design for a given application.

Figure 4b shows a section of the known-subset resolved on a 2D space. Despite the complexity of the optimization problem, the objective function is not as irregular as in the previous example (Figure 3c), which suggests that algorithms might act well on such a search space. Nevertheless this objective function is more complex than the case studied Figure 2b, and hence many more objective function evaluations may be needed to find optimal results.

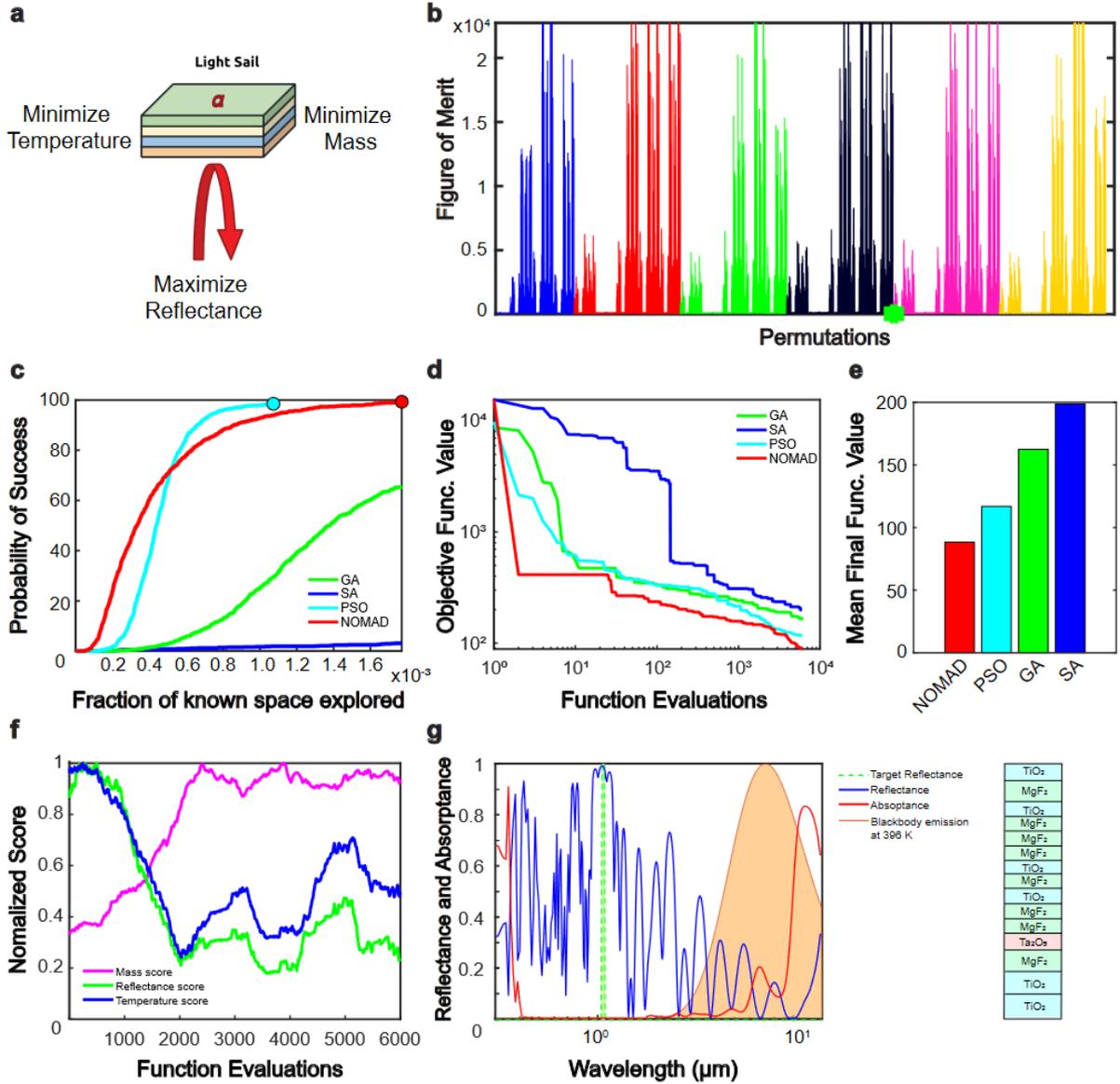

*Figure 4: (a) Schematic Illustration of the optimization problem. The objective is to minimize lightsail mass, maximize laser reflectivity across 1050nm – 1080nm wavelength window, and minimize lightsail temperature. (b) One-dimensional graph of the objective function designed on the 8-dimentsional known-subset. A small fraction of the permutations on the known-subset is shown. Each point corresponds to a different multilayer structure, whereas different colors represent different material permutations. A marker point denotes a global optimum of the known-subset. (c) Probability of success in finding the global optimum for fixed-target search on a known-subset for different algorithms. (d) Fixed-cost performance analysis. Average best FoM found by the algorithms is shown. (e) Bar-chart depicts the average best FoM found by the algorithms after 6000 function evaluations. (f) A rolling average of lightsail temperature, reflectance and mass metrics as by one of NOMAD run. (g) Reflectance and absorptance spectra corresponding to the best result found by NOMAD after 25,000 function evaluations. Inset provides a schematic illustration of structure found.*

Figure 4c plots the results of fixed-target analysis on the known-subset (Eq. (4)). We observe that NOMAD and PSO are able to find global optimum efficiently, as expected for structured objective functions. NOMAD finds global optimum more often than PSO, however at a slightly higher evaluation cost. GA does go on to find the global minimum reliably, but at an average evaluation cost nearly double that of PSO and NOMAD. Results of fixed-cost analysis on the entire search space are plotted in Figure 4d. We find that NOMAD consistently outperforms other algorithms in finding optimal solutions. Hence, the average best value found by NOMAD after 6000 function evaluations is ~45% lower than that of PSO and is twice better than that of GA, Figure 4e.

It is instructive to examine how the temperature, mass and reflectance individually evolve during the process of optimization. Figure 4f shows a running average value of these functions for one of the NOMAD's trials. Initially starting with a random selection of materials and layer thicknesses temperature of the structures is very high and the reflectivity is very low. As the algorithm explores the search space the temperature rapidly decreases while reflectance increases. However, the performance gained on these two metrics (i.e., temperature and reflectance) comes at the cost of an increased mass. Although NOMAD delivers an absolute best result, as compared to other algorithms, our choice of a very complex search space with 15 material choices results in finding designs that are far from optimal even after many function evaluations. Indeed, after 6000 function evaluations the best result found by NOMAD is a structure with the following material combinations: *{$TiO_2$, $MgF_2$, $TiO_2$, $MgF_2$, $Si_3N_4$, $MgF_2$, $MgF_2$, Au, Cu, $MgF_2$, Al, Al, $MgF_2$, Al, $MgF_2$}*. The structure presents an alternating combination of high and low index materials with a metal backreflector, which allows achieving 0.9992 reflectance in the spectral band of interest. However, presence of metals results in a very high temperature $T \simeq 3800\ K$ (our models do not account for change of refractive index with temperature and materials melting). With 25000 balckbox function evaluations NOMAD identifies all dielectric design with an alternating combination of high and low refractive index layers: *{$TiO_2$, $MgF_2$, $TiO_2$, $MgF_2$, $MgF_2$, $MgF_2$, $TiO_2$, $MgF_2$, $TiO_2$, $MgF_2$, $MgF_2$, $Ta_2O_5$, $MgF_2$, $TiO_2$, $TiO_2$}*, which is enables creating a photonic bandgap in range of the wavelengths of interest (reflectance ~0.99). Low absorbance of the dielectric materials and emissivity in the near-infrared due to polar resonances of $TiO_2$ allow keeping sail temperature below 400K. Corresponding spectra of the multilayer structure after 25000 evaluations are shown in Figure 4g. This example clearly illustrates that while algorithms may be tuned to escape local minima and efficiently traverse the search space choice of complex and ill-defined search spaces may lead to suboptimal solutions even after extensive search procedures applied. Preselecting material options would have allowed better initial conditions and faster algorithm convergence.

**Table 1. Summary of algorithm performance on structured search spaces.**

| Algorithm | Search method | Reliability | Efficiency |
| --- | --- | --- | --- |
| NOMAD | Direct search with metaheuristic | ~100% | High |
| PSO | Heuristic with strong search mechanic | >95% | High |
| GA | Fully heuristic | ~100% | Moderate |
| SA | Fully heuristic | ~70% | Low |

**Conclusions**

Our results show several important conclusions that will help formulating optimization problems in more complex inverse design scenarios. Firstly, we find that all algorithms irrespective of their search engine operate efficiently on objective functions that exhibit some structure. Therefore, defining FoM yielding structured objective functions will dramatically expedite discovery of optimal solutions. Secondly, we find that direct search methods with metaheuristics (e.g., NOMAD) are more efficient than heuristic algorithms (e.g., GA) in discovering optimal solutions for structured objective functions. In Table 1 we provide a corresponding summary of algorithm performance for structured objective functions. Probability of finding global optimum on known-subset and efficiency in finding optimal solutions in the least function evaluations are shown. Thirdly, our results indicate that the trends in small subspaces map to the structure of larger spaces, which allows utilizing smaller subspaces to explore and tune algorithms prior to searching on larger computationally intractable search spaces. Lastly, we underline the importance of preselecting the search space and identifying initial conditions. Relying even on very efficient algorithms to find optimal solutions on ill-defined optimization problems is likely to result in solutions that are far from optimal. Indeed, as our analysis of know-subsets shows algorithms require traversing ~0.1% of the search space to find the optimal solution reliably. To conclude we have developed and explored a framework that allows better understanding algorithm performance for photonic inverse design. Our framework and results can act as valuable utilities for tackling problems that require structures with performance close to the global optimum.

## Methods

Our spectral discretization in the interval [0.3 $\mu m$ 14 $\mu m$] is as follows: we use 10 $nm$ step from 0.3 $\mu m$ to 10 $\mu m$ and 100 $nm$ step from 10 $\mu m$ to 14 $\mu m$, which leads to a total of 1012 spectral points.

For the sake of computational simplicity single sided emissivity was considered for all optimization examples. We do not expect our conclusions on algorithm performance to change with accounting double sided emissivity. The refractive index data to materials examined in the paper was taken from the refractiveindex.info database. Minimum value of material imaginary permittivity was set at $10^{-7}$ where appropriate.

We tuned the optimization algorithms to ensure convergence for objective functions examined in this paper. We used the GERAD implementation of NOMAD [36, 59] and we found that enabling the Variable Neighborhood Search (VNS) metaheuristic to complement NOMAD's direct search strategy [37] helped drastically improve performance by helping it escape local minima. Latin Hypercube sampling was implemented. For the remaining algorithms, we used the implementations provided in MATLAB's Global Optimization Toolbox. Population size for GA is 200. Swarm size in PSO was 100. We tuned their stopping conditions according to the scaling of the objective function. For SA, we added more emphasis on random selection of points as it explores the search space by tuning its acceptance function so that it was better able to handle the more chaotic objective functions.

## Author Contributions
The manuscript was written through contributions of all authors.

## Notes
The authors declare no competing financial interest.

## Acknowledgments
Authors acknowledge useful discussions with Kostia Zuev and Katherine Fountain, and thank Haonan Ling for help with writing the manuscript.